\documentclass[journal,twoside]{IEEEtran}
\ifCLASSINFOpdf
\else
\fi
\usepackage{xfrac}
\usepackage{mathtools,amsmath}
\usepackage{multirow}
\usepackage{gensymb}
\usepackage[x11names,table]{xcolor}
\usepackage{amssymb}
\usepackage[permil]{overpic}
\usepackage{pict2e} 
\usepackage{cite}

\usepackage{hyperref}
\usepackage{lipsum}
\usepackage{algpseudocode,algorithm}
\DeclareMathAlphabet{\mathpzc}{OT1}{pzc}{m}{it}
\usepackage[cal=dutchcal]{mathalfa}
\ifCLASSOPTIONcompsoc
\usepackage[caption=false,font=normalsize,labelfon
t=sf,textfont=sf]{subfig}
\else
\usepackage[caption=false,font=footnotesize]{subfi
g}
\fi

\definecolor{mycolor}{RGB}{218,232,252}
\rowcolors{2}{mycolor!60}{cyan!25}

 \usepackage{etoolbox}

\usepackage{stfloats}

\makeatletter
\patchcmd{\@maketitle}
  {\addvspace{0.5\baselineskip}\egroup}
  {\addvspace{-1\baselineskip}\egroup}
  {}
  {}
\makeatother

\begin{document}
\bstctlcite{IEEEexample:BSTcontrol}


\title{Reconfigurable Intelligent Surfaces in Upper Mid-Band 6G Networks: Gain or Pain?}

%
%
%

\author{ Ferdi Kara,~\IEEEmembership{Senior Member,~IEEE,} \"Ozlem Tu\u{g}fe Demir,~\IEEEmembership{Member,~IEEE,} Emil Bj\"ornson~\IEEEmembership{Fellow,~IEEE.}  
\thanks{F. Kara is with the KTH Royal Institute of Technology, Sweden, and also with Zonguldak B\"ulent Ecevit University, T\"urkiye. \"O. T. Demir is with  TOBB University of Economics and Technology, T\"urkiye. E. Bj\"ornson is with the KTH Royal Institute of Technology, Sweden. The work of F. Kara and E. Bj\"ornson is supported by the Swedish Foundation for Strategic Research and the SweWIN Vinnova Competence Center. The work of \"O. T. Demir is supported by 2232-B International Fellowship for Early Stage Researchers Programme funded by the Scientific and Technological Research Council of T\"urkiye.}}

\maketitle
\begin{abstract}
Reconfigurable intelligent surfaces (RISs) have emerged as one of the most studied topics in recent years, hailed as a transformative technology with the potential to revolutionize future wireless systems. While RISs are recognized for their ability to enhance spectral efficiency, coverage, and the reliability of wireless channels, several challenges remain. Notably, convincing and profitable use cases must be developed before widespread commercial deployment can be realized.  The first sixth-generation (6G) networks will most likely utilize upper mid-band frequencies (i.e., 7-24\,GHz). This is regarded as the \textit{golden band} since it combines good coverage, much new spectrum, and enables many antennas in compact form factors. There has been much prior work on channel modeling, coexistence, and possible implementation scenarios for these bands. 
There are significant frequency-specific challenges related to RIS deployment, use cases, number of required elements, channel estimation, and control. These are previously unaddressed for the upper mid-band.
In this paper, we aim to bridge this gap by exploring various use cases, including RIS-assisted fixed wireless access (FWA), enhanced capacity in mobile communications, and increased reliability at the cell edge. We identify the conditions under which RIS can provide major benefits and optimal strategies for deploying RIS to enhance the performance of 6G upper mid-band communication systems.
\end{abstract}


The rapid societal adoption of wireless communication technologies has led to increased demand for higher data rates, enhanced reliability, and greater network capacity \cite{ITU2023}. As we move towards the 6G era, the upper mid-band frequency range, spanning from 7\,GHz to 24\,GHz, has emerged as the critical spectrum to meet these demands \cite{Kang2024cellular}. The upper mid-band offers a unique balance between coverage and capacity, with higher bandwidth availability than in the sub-7 GHz bands and more favorable propagation characteristics than in millimeter-wave bands. This makes it an attractive candidate for 6G applications, offering a promising compromise between the extensive coverage of lower bands and the high data rates of higher bands. To this end, after careful consideration at the World Radiocommunication Conference 2023 (WRC-23), the International Telecommunications Union (ITU) identified two candidate segments of the upper mid-band for potential 6G use: 7.125-8.4\,GHz and 14.8-15.35\,GHz \cite{WRC23}. The first 6G deployments will most likely be in one of these bands.

Over the past five years, reconfigurable intelligent surfaces (RISs) have gained significant attention as a transformative technology that enables control of wireless propagation environments. An RIS is composed of sub-wavelength-sized elements capable of dynamically modifying the electromagnetic properties of incident waves before reflecting them. By controlling the phase, amplitude, and polarization of the incident waves, an RIS can enhance signal quality and coverage of communication links in real-time, as well as improve other properties such as channel rank and interference \cite{RIS_principles}. Some researchers describe RIS as a game-changer for the next generation of wireless systems \cite{RIS_rui_zhang}, while others describe it as a new tool whose practical use case is still to be discovered \cite{RIS_myths_magazine}. A well-configured RIS can improve any wireless technology, including massive multiple-input multiple-output (MIMO) systems, integrated sensing and communications, and terahertz communications, but there are also competing ways to enhance networks \cite{RIS_myths_magazine}.
Nevertheless, the technology readiness level of RIS has evolved rapidly \cite{RIS_experimental}. 
Although the potential of RIS has been explored excessively, the theoretical studies are mostly frequency-agnostic, while simulations and measurements are done in the sub-7 GHz band. However, there are significant frequency-dependent challenges in RIS deployment. The upper mid-band
remains unexplored in the RIS context, but is an essential research avenue with significant potential to affect actual 6G deployments. Moreover, some previous RIS studies overlook some specific benefits of RIS or ignore some realistic conditions (e.g., the static path between source and destination is commonly neglected). Therefore, its advantages are sometimes overemphasized and a true benchmark comparison is missing. In this context, RIS deployments should be further evaluated with practical constraints.

In this paper, we take major steps towards closing this gap by examining the potential use of RIS in 6G upper mid-band networks. We investigate several use cases, evaluating each scenario based on either spectral efficiency (SE) or coverage probability, as well as the required number of RIS elements for effective RIS assistance. In this way, we aim to determine the conditions under which the RIS can offer significant improvements that can merit real-world deployment and the best strategies for deploying RIS in the upper mid-band. Through this study, we have identified four scenarios where RIS can enhance 6G networks in this way. Nonetheless, several challenges need to be tackled before any commercial rollouts should be considered. We discuss these challenges and outline the open questions before concluding the paper. Our results aim to guide future research and development, paving the way for the successful implementation of RIS-enabled upper mid-band solutions in 6G.

\begin{figure*}[!t]
\centering
\fcolorbox{mycolor}{mycolor}{
\subfloat[$f_c=7.8$ GHz]
{ \includegraphics[width=0.33\textwidth]{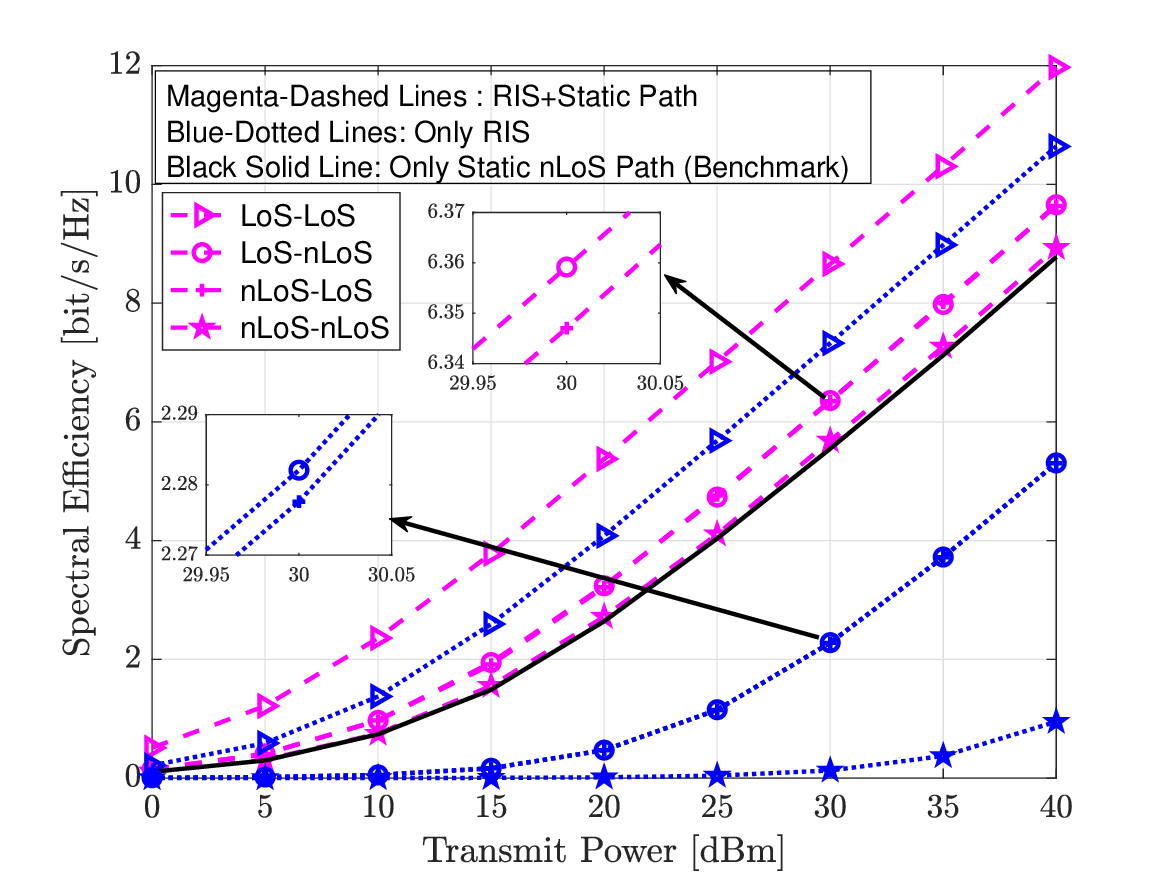}
}
\centering
\subfloat[$f_c=15$ GHz]
{ \includegraphics[width=0.33\textwidth]{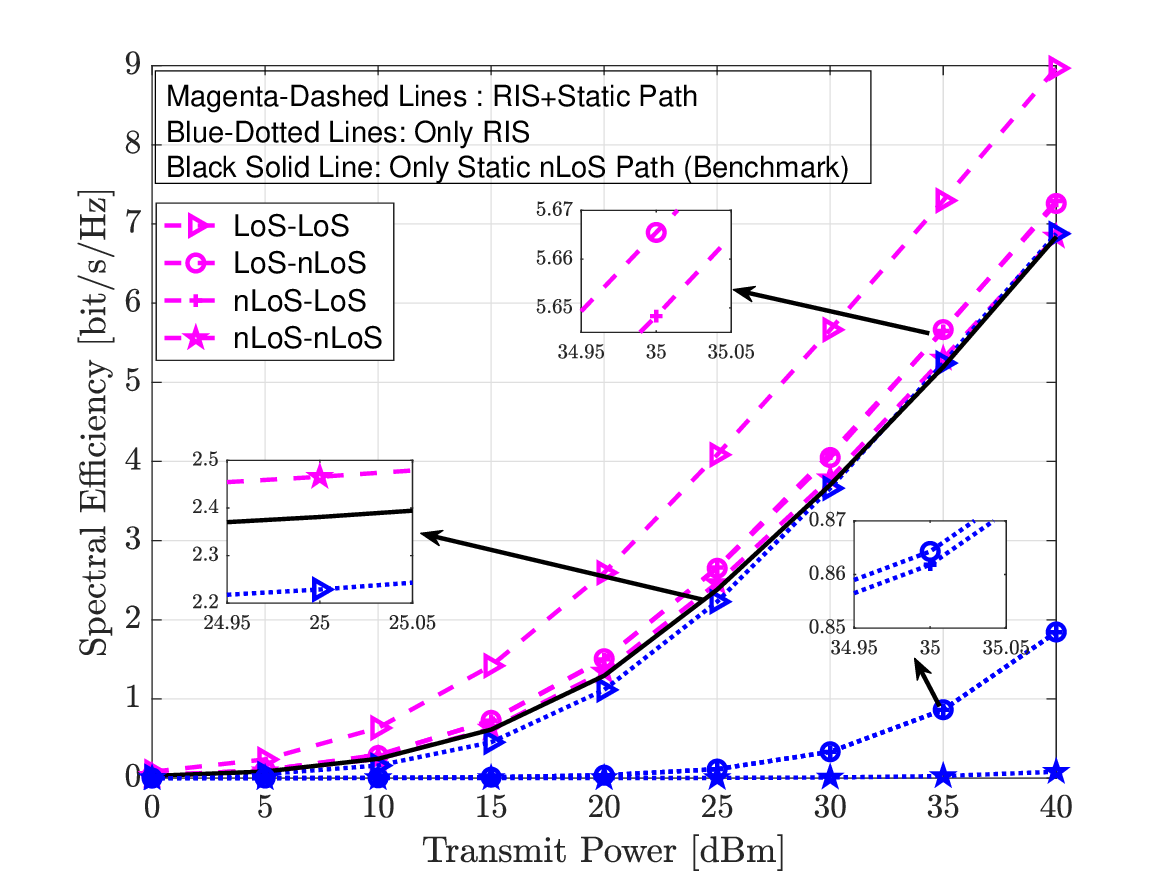}}
\centering
\subfloat[Use Case 1: RIS-assisted FWA]
{ \includegraphics[width=0.33\textwidth]{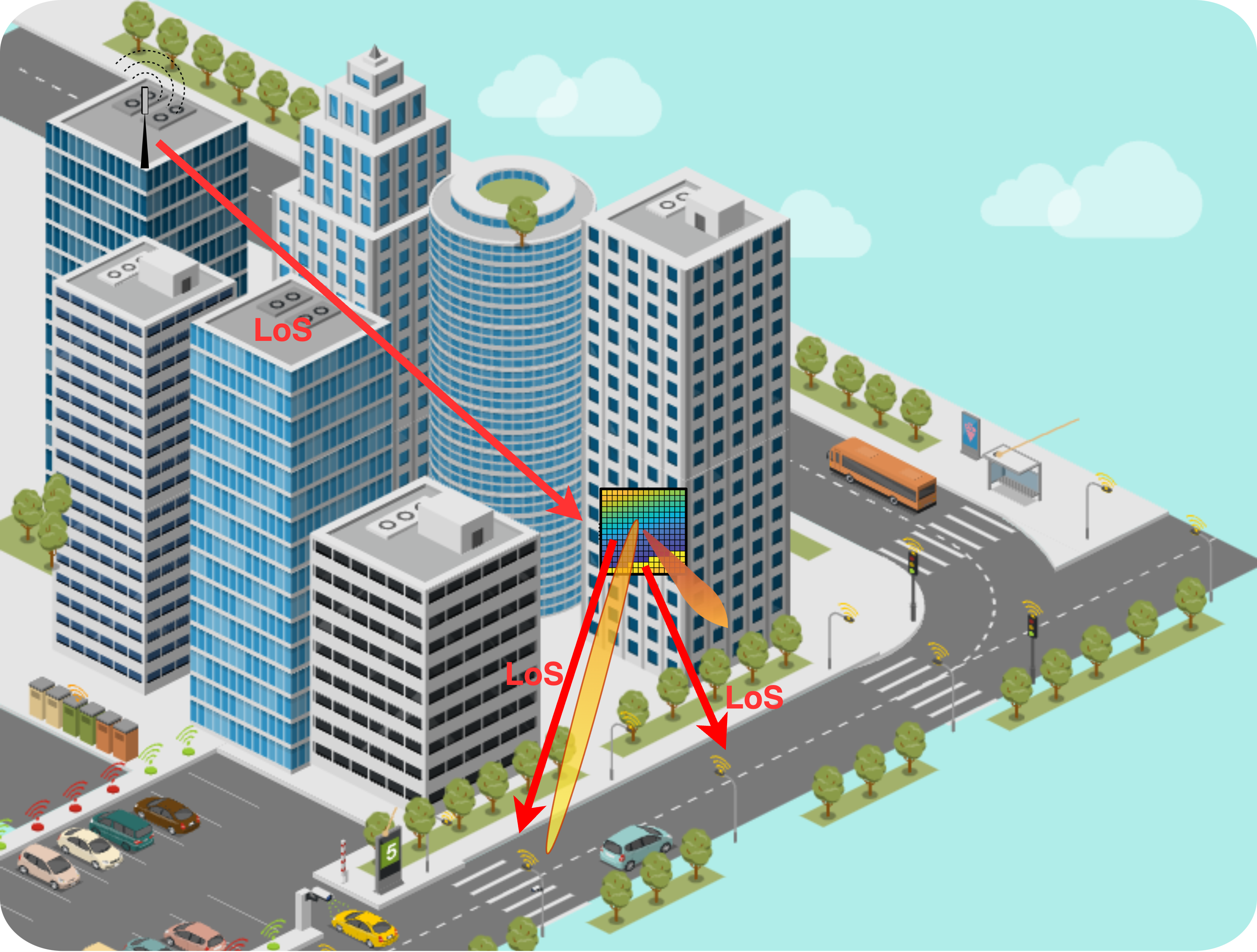}
}
}
\caption{ SE performance comparisons for RIS-assisted communications. Node locations: transmitter (0,0)\,m, receiver (0, 100)\,m, RIS (25, 50)\,m. In the legends, pairs refer to transmitter-RIS and RIS-receiver links. The transmitter-receiver link is considered to be nLoS. Accordingly, the first beneficial scenario is Use Case 1: RIS-Assisted Fixed Wireless Access (FWA) when both Tx-RIS and RIS-Rx links have LoS.}
\label{fig:various_LoS_scenarios}
\end{figure*}

\section*{When are RISs Beneficial? Possible Use Cases}

In this section, we aim to identify potential use cases for RIS-assisted communication in networks operating in the upper mid-band. To this end, we must first address two key questions: 1) When does an RIS provide significant benefits?; 2) Where should the RIS be deployed? By answering these questions, we can better elucidate the advantages of RIS. The following subsections outline our perspectives on the four potential RIS deployment scenarios based on the responses to these questions. These use cases are explained and evaluated in detail in the following sections. 

\subsection*{RIS-Assisted Fixed Wireless Access}

In much of the RIS literature, it is assumed that the static path
between the transmitter and receiver does not exist, allowing the RIS to create a previously non-existing communication link between these nodes. Is this a valid assumption, or does it lead to biased results that overemphasize the benefits of RIS-assisted communications? In a wireless communication environment, radio-frequency (RF) waves reach their destination through various propagation effects. Consequently, even if a line-of-sight (LoS) path is absent, a static non-LoS (nLOS) path is often available between the transmitter and receiver through diffuse/specular reflections, diffraction, or transmission through objects. In fact, if there exists a surface on which an RIS could be deployed to establish a ``new'' reflection path to the receiver, it would be strange if that surface did not already reflect some small amount of RF energy toward the receiver. It is mostly at 
sub-THz or optical band communication that the reflection and absorption properties are so unfavorable that RF waves can only make it to the receiver through LoS paths, or ``virtual'' LoS paths created by an RIS.
In the upper mid-band, however, rich multi-path effects still occur, meaning that nLoS static paths (and possibly an LoS path in some cases) will be available for receivers within the coverage area \cite{measurement_industrial}. Therefore,
the RIS deployment does not establish new links but merely enhances the performance of point-to-point links that already exist.

\begin{table}[!t]
    \centering
        \caption{System Setup and Simulation Parameters}
    \begin{tabular}{c||c}
        \rowcolor{cyan!60} \textbf{Parameter}& \textbf{Value} \\ \hline \hline
        Carrier frequency ($f_c$)  (GHz) & 7.8, 15 \\ \hline
        Bandwidth (MHz) &400 \\ \hline
        Transmit power (dBm) & [0:50] \\ \hline
        Noise spectral density (dBm/Hz) & -174 \\ \hline
        & 3GPP UMi \cite{ETSI} \\
        \cellcolor{cyan!25}& \cellcolor{cyan!25}Probabilistic LoS \\
        \multirow{-3}{*}{\cellcolor{cyan!25}{Channel model} }&Far-field channel characteristics\\ 
        \hline
         Number of channel realizations  & $10^5$\\ \hline 
         Channel state information & Perfect cascaded phase information\\ \hline
        Number of elements at RIS ($N$) & 2000 \\ \hline
         & Optimal phase shifts\\
        \multirow{-2}{*}{\cellcolor{cyan!25}{Phase shifts at RIS}}&\cellcolor{cyan!25}Infinite resolution \\ \hline
         \cellcolor{mycolor!60}& \cellcolor{mycolor!60} Transmitter \\  & \cellcolor{mycolor!60} $G_T=10$  \\ 
      \cellcolor{mycolor!60} & \cellcolor{mycolor!40} Receiver \\  \multirow{-4}{*}{\shortstack{Directivity (dBi) \\ (Antenna/beamforming gains)}} & \cellcolor{mycolor!40}$G_R=3$ \\ \hline
        Transmitter height (m) & 10 \\ \hline
         & Use Case 1 \\ \cellcolor{mycolor!60} & \cellcolor{mycolor!60} 2.5 \\ 
        & \cellcolor{mycolor!40}  Use Cases 2, 3, 4  \\ \multirow{-4}{*}{\cellcolor{mycolor!60} User height (m)} & \cellcolor{mycolor!40} 1.5 \\ \hline
       \cellcolor{cyan!25} RIS height (middle point) (m) & \cellcolor{cyan!25}5 \\ \hline
       \cellcolor{mycolor!60} Cell radius (m) &\cellcolor{mycolor!60} 200 \\ \hline
       \cellcolor{cyan!25}Transmitter location $(x,y)$ &\cellcolor{cyan!25} $(0, 0)$\,m\\ \hline
        \cellcolor{mycolor!60} & \cellcolor{mycolor!60}Use Case 1 \\ & \cellcolor{mycolor!60}$(0,100)$\,m \\  \cellcolor{mycolor!60}& \cellcolor{mycolor!40}Use Case 2 \\ & \cellcolor{mycolor!40}2000 users \\ \cellcolor{mycolor!60}&\cellcolor{mycolor!40}(Randomly deployed in the cell) \\  & \cellcolor{mycolor!60} Use Case 3 \\\cellcolor{mycolor!60} &  \cellcolor{mycolor!60} 200 users \\ & \cellcolor{mycolor!60}(Randomly deployed in the ROI) \\\cellcolor{mycolor!60}  & \cellcolor{mycolor!40}Use Case 4 \\
       \multirow{-10}{*}{\cellcolor{mycolor!60}{\shortstack{Receiver(s) location $(x,y)$ \\ (or definition)}}}& \cellcolor{mycolor!40}Cell edge $(0,200)$\,m \\ \hline
        & Use Case 1 \\  \cellcolor{cyan!25}&\cellcolor{cyan!25}$(25,50)$\,m \\  & \cellcolor{cyan!15}Use Case 2 \\ \cellcolor{cyan!25} & \cellcolor{cyan!15}LoS-always distance to transmitter \\ & \cellcolor{cyan!15}($<18$\,m) \\ \cellcolor{cyan!25} & \cellcolor{cyan!25}Use Case 3 \\  & $(100,100)$\,m \\ \cellcolor{cyan!25} & \cellcolor{cyan!15}Use Case 4 \\ &\cellcolor{cyan!15}LoS-always distance to cell edge \\ \multirow{-10}{*}{\cellcolor{cyan!25}\shortstack{RIS location $(x,y)$ \\ (or definition)}} &\cellcolor{cyan!15}($<18$\,m) \\ \hline
    \end{tabular}
    \label{table:simulation_prameters}
\end{table}

Fig.~\ref{fig:various_LoS_scenarios}.a and b illustrate the SE of an RIS-assisted communication system operating in the upper mid-band. The common parameters for all simulations throughout the paper are provided in Table \ref{table:simulation_prameters}. Throughout the paper, in the simulations, we present results for three different scenarios: \textit{i)} RIS-assisted communication with a static path; \textit{ii)} RIS-enabled communication with no static path between transmitter and receiver (a common scenario in the prior literature); and \textit{iii)} a benchmark with only a static nLoS path (i.e., point-to-point communication). Besides, the multi-antenna configurations for both ends are mimicked by antenna/beamforming gain as given in Table \ref{table:simulation_prameters} where the stated values are averaged from industrial reports to reflect real-world applications. In Fig.\ref{fig:various_LoS_scenarios}, for the RIS-assisted scenarios, we also switch between having LoS and nLoS conditions for the paths between the transmitter/receiver and RIS with realistic channel models from \cite{ETSI} that consider frequency-dependent characteristics and support performance evaluation in the upper mid-band.

From this figure, we observe that RIS-assisted communication can yield significant improvements when a static path is available, but not necessarily. The performance gains are subtle unless both the transmitter-RIS and RIS-receiver links have LoS conditions. 
Similarly, RIS-assisted communication without a static path only outperforms point-to-point communication (with a static path) if LoS paths are ensured to and from the RIS. Nevertheless, one might argue that even in this scenario, the signal-to-noise ratio (SNR) gain over point-to-point communication is less than 3-5\,dB, so the same SE can be achieved by simply doubling transmit power or with more directive beamforming rather than introducing an RIS with all its associated deployment and control complexities. We notice that this minor gain disappears when a higher carrier frequency (e.g., 15\,GHz) is deployed. This indicates that even $N=2000$ RIS elements (in the absence of a static path) are insufficient to surpass point-to-point communication at this frequency and communication range. The higher frequency means worse path loss conditions per antenna/element and it requires more elements on RIS to compensate for the cascaded (e.g., multiplied) fading distortions. This will be further discussed in the following sections.

The results in Fig.~\ref{fig:various_LoS_scenarios}.a and b demonstrate that when considering the true benchmark (with a weak but not neglected static path), RIS can only provide meaningful improvements in specific scenarios. This brings us back to the question: Are RISs truly beneficial or is the hype created by unrealistic propagation assumptions?
The simple answer is: No, it is not hype, but RISs can be a game-changer in certain scenarios. Fig.~\ref{fig:various_LoS_scenarios} shows that RIS-assisted communication provides up to 35-55\% higher SE than point-to-point communications. The only concern here is that the RIS must be deployed with LoS path to the base station and can only assist users that it also has LoS paths to. 
This leads us to \textbf{Use Case 1: RIS-Assisted Fixed Wireless Access (FWA)}, as illustrated in Fig.\ref{fig:various_LoS_scenarios}.c, where the user device is a fixed access point (router) for a personal network, smart-city surveillance systems, etc., that connects to a fixed base station. 5G FWA is already a commercial success in some countries with low penetration of fiber connections. The capacity and number of customers per base station can be vastly increased in the upper mid-band by deploying RISs that extend the coverage by creating virtual LoS paths. 
This is relatively easy to achieve in FWA scenarios since the transmitter and receiver positions are fixed. 
Ray-tracing might be sufficient for network operators to predict the extended coverage, and the location of the user equipment (UE) can be fine-tuned to ensure LoS to the RIS.

\begin{figure*}[!t]
\centering
\fcolorbox{mycolor}{mycolor}{
\subfloat[SE and E2E LoS vs RIS location]{\begin{overpic}[width=0.315\textwidth]{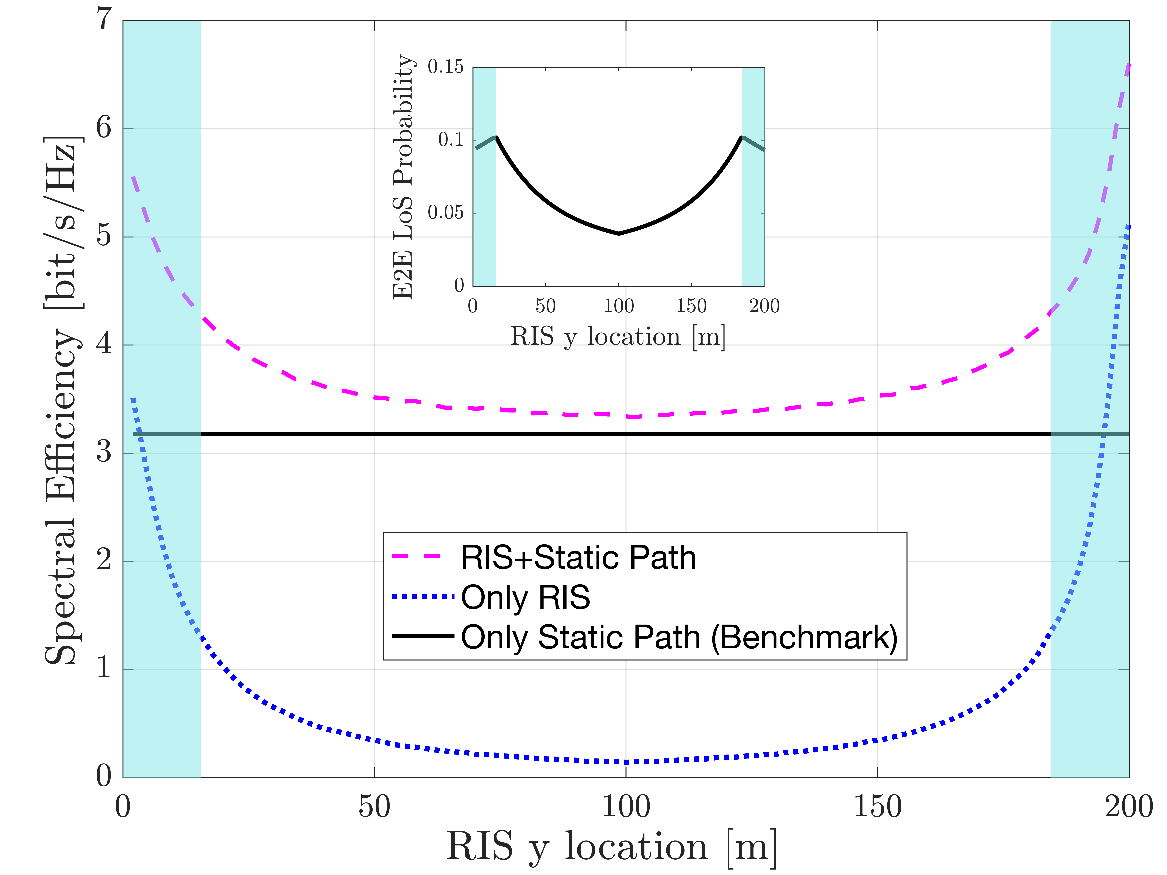}%
\put(170,800){\color{red}\textbf{LoS Guaranteed Regions}}%
\put(40,720){\color{red}(\textbf{Transmitter-RIS)}}%
\put(200,710){\linethickness{.5mm}\color{red}\vector(-0.7,-1){60}}
\put(600,720){\color{red}(\textbf{RIS-Receiver)}}%
\put(870,710){\linethickness{.5mm}\color{red}\vector(0.7,-1){60}}
\end{overpic}
}
\centering
\subfloat[Use Case 2]{\includegraphics[width=0.33\textwidth]{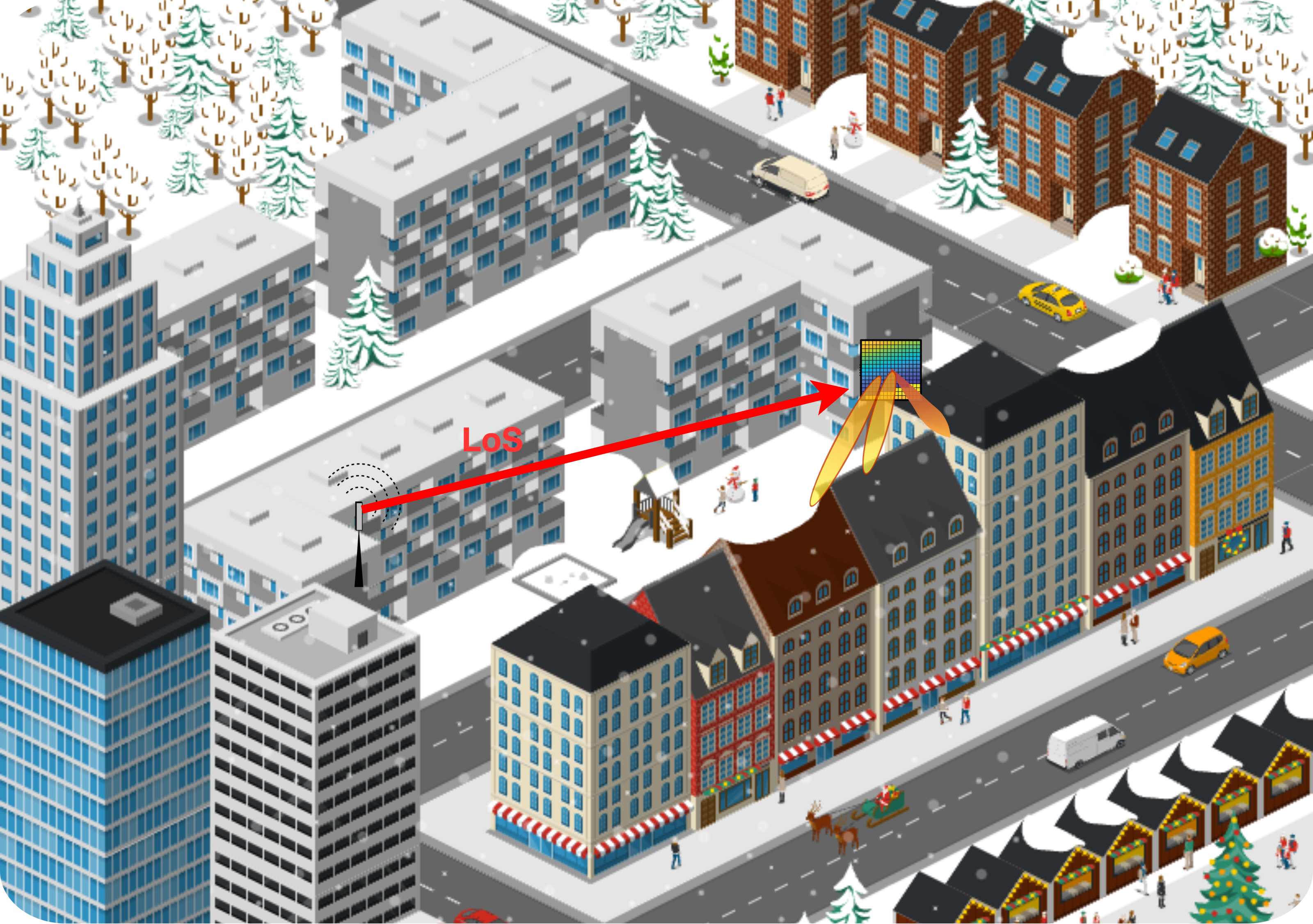}
}
\centering
\subfloat[Use Case 3]{\includegraphics[width=0.33\textwidth, height=4.7cm]{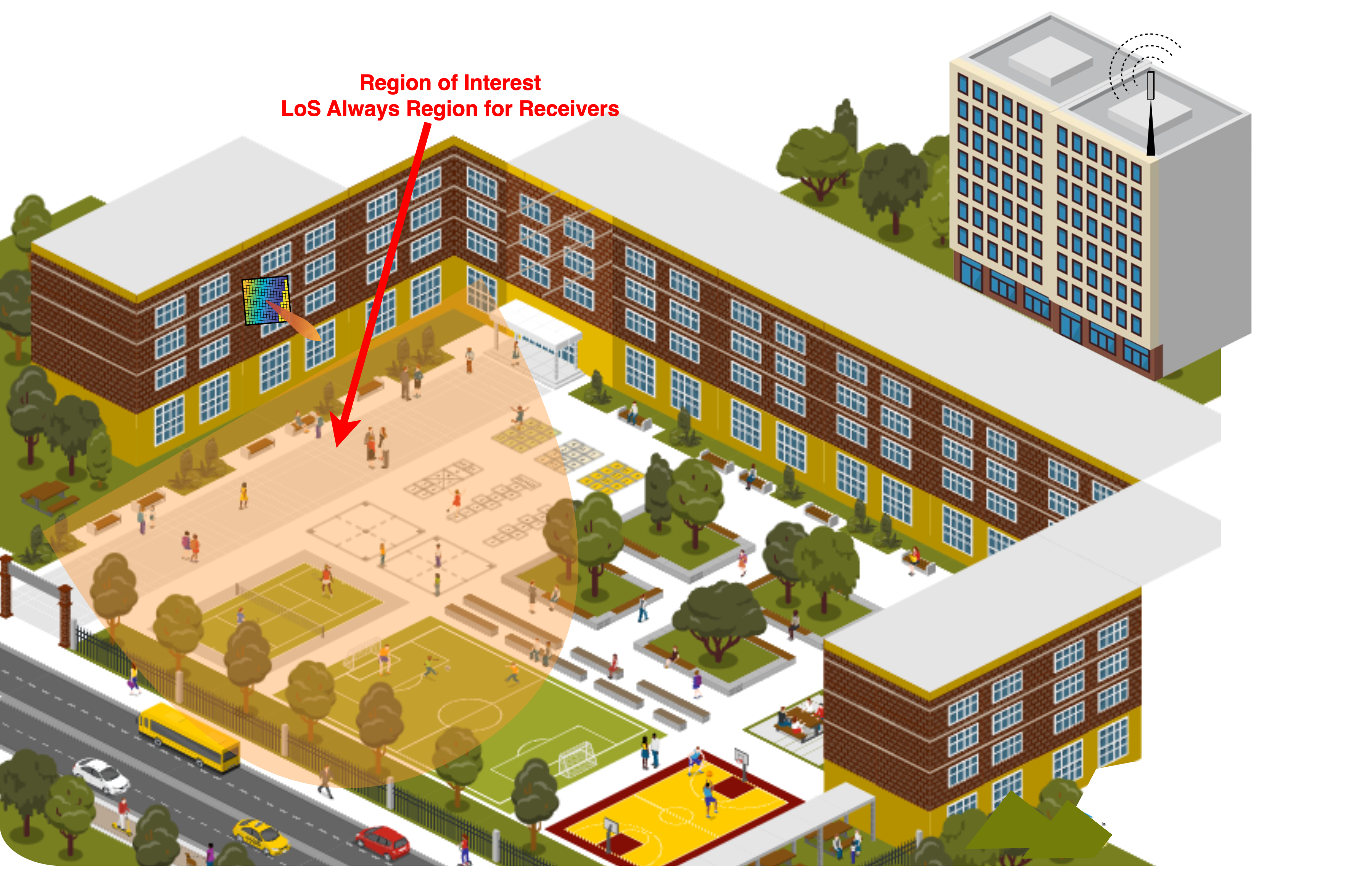}}
}
\caption{SE performance comparisons for RIS-assisted communications vs RIS location at $f_c=7.8$\,GHz with a realistic/probabilistic LoS channel model (e.g., 3GPP UMi\cite{ETSI}). Node locations ($x,y,z$): transmitter (0,0,10)\,m, receiver (0, 200, 2)\,m, RIS ($\mathrm{RIS}_x$, $\mathrm{RIS}_y$, 5)\,m. According to the $y$-axis location of RIS ($\mathrm{RIS}_y$), we always pick an $x$-axis location ($\mathrm{RIS}_x$) to have 30$^{\circ}$ azimuth angle between transmitter-RIS or RIS-receiver, whichever RIS is closed to. Transmit power is $T_X=30$\,dBm. Accordingly, there are two possible use cases to enhance mobile capacity. Use Case 2: Enhanced Capacity For Distributed Users within the Cell when RIS is in the LoS region of the transmitter. Use Case 3: Enhanced Capacity for a Region of Interest (ROI) when RIS has LoS path for all users in ROI.}
\label{fig:RIS_placements}
\end{figure*}

\subsection*{Enhanced Capacity in Mobile Communication within the Cell }

In the previous subsection, we demonstrated that LoS paths are essential for achieving performance gains in RIS-assisted communication. Such conditions are harder to guarantee for mobile networks, where UE locations are dispersed both temporally and spatially. Consequently, establishing a LoS path is much more challenging than in FWA, as it is influenced not only by the positions of transceivers but also by surrounding objects and their movements. Hence, the RIS node must be strategically placed to ensure a LoS path can be formed at both ends. Is this always feasible? In urban mobile communications, the answer is generally no unless the transmitter, RIS, and receiver nodes are nearby (i.e., within a few meters) of each other. In mobile scenarios, the probability of LoS \cite{ETSI} should be factored into the design, necessitating careful placement of the RIS.

To substantiate these claims, Fig.~\ref{fig:RIS_placements}.a shows the SE of RIS-assisted communication relative to the RIS location, considering the LoS probability in the upper mid-band frequency of $7.8$\,GHz. It is evident that unless a LoS path is assured (as in FWA), positioning the RIS midway between the transmitter and receiver is the least favorable scenario. The SE can drop to nearly zero when only RIS-assisted communication is available without a static path. Nevertheless, there are many papers studying midway RIS deployments because that is optimal for traditional relay deployments.

\begin{figure*}[!t]
\centering
\fcolorbox{mycolor}{mycolor}{
\subfloat[in Use Case 2]{\includegraphics[width=0.43\textwidth]{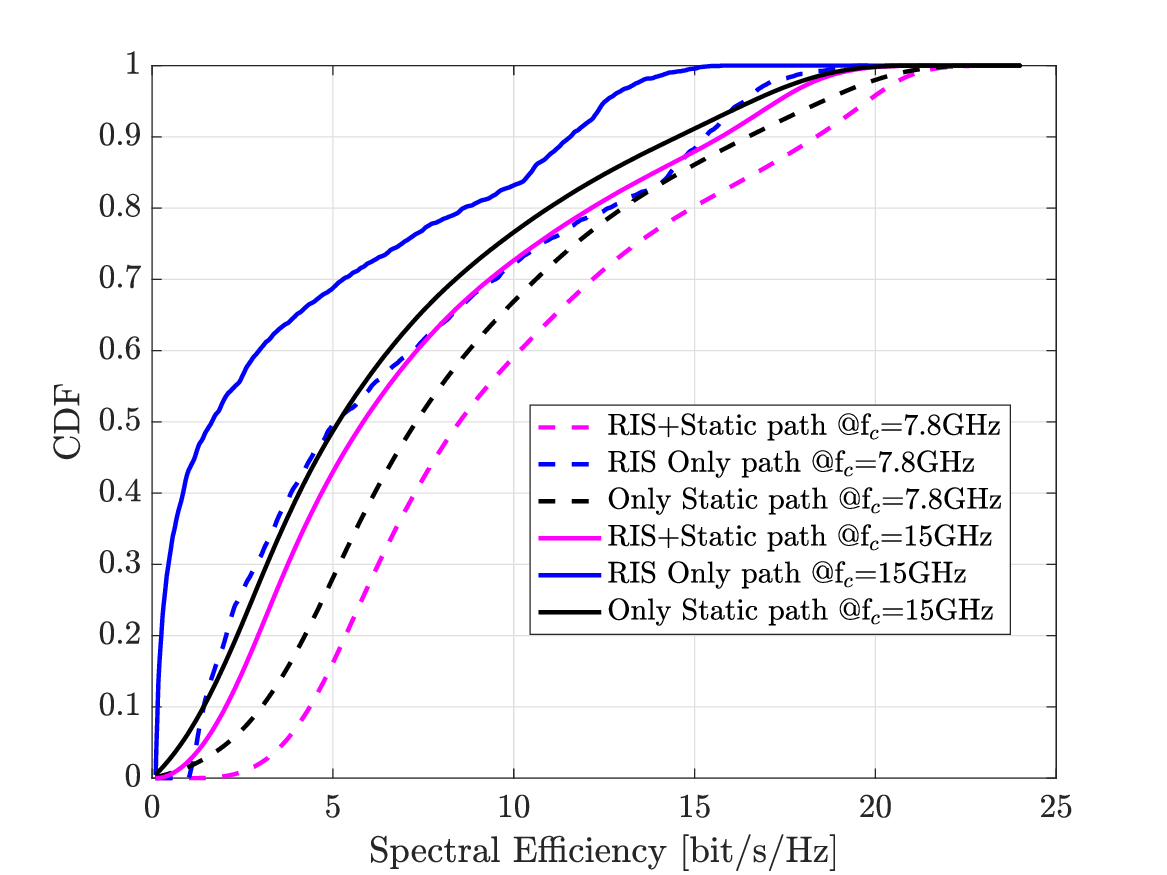}
}
\subfloat[in Use Case 3]{\includegraphics[width=0.43\textwidth]{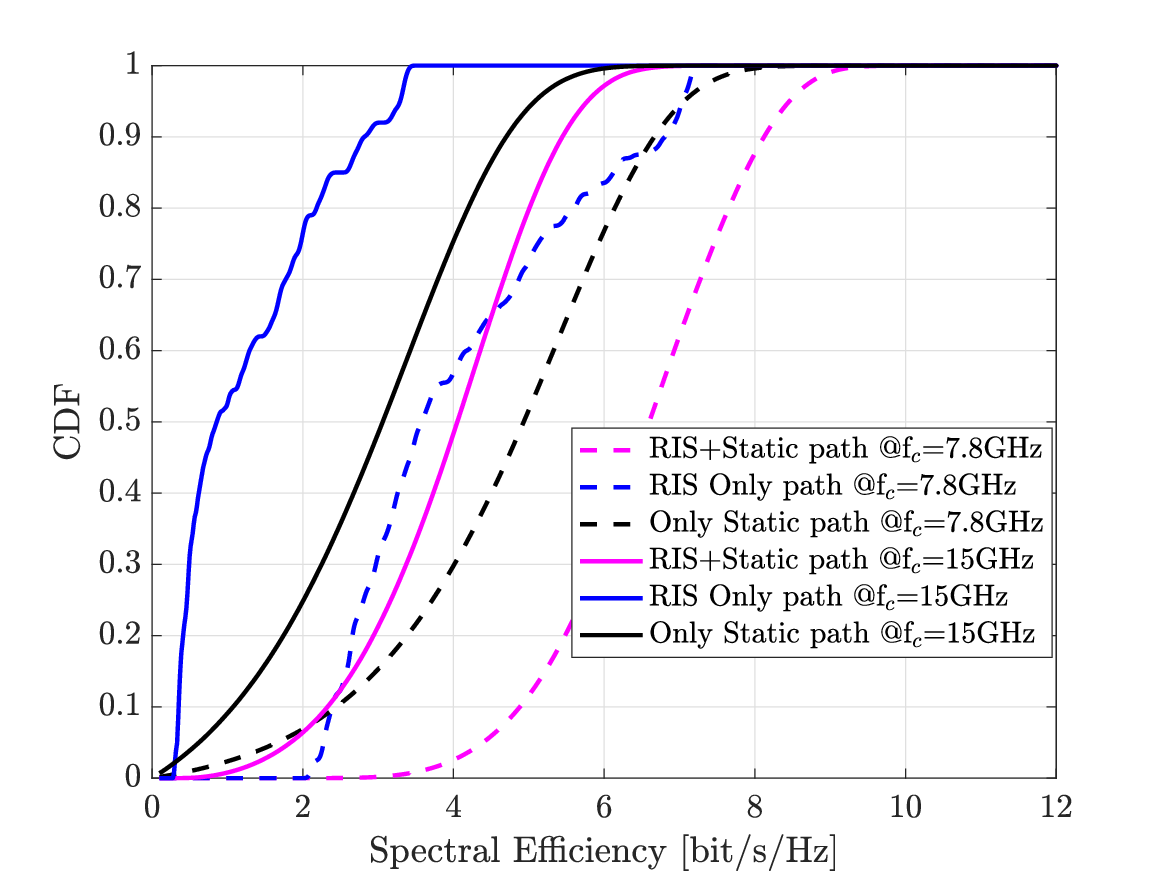}
}
}%
\caption{The CDF of the SE at $T_X=30$\,dBm. The left-hand side figure shows  Use Case 2 where the SE is achieved at different 2000 random user locations within the cell and RIS is located in the LoS region of the transmitter. The right-hand-side figure shows  Use Case 3 where the SE is achieved at different 200 random user locations within the region of interest where RIS has LoS links for all users in this region.}
\label{fig:cdf_vs_rate_incell}
\end{figure*}

Fig.~\ref{fig:RIS_placements}.a also shows that the end-to-end (E2E) LoS probability for the cascaded transmitter-RIS-receiver path is very low, and an E2E LoS path can never be guaranteed in mobile scenarios. Consequently,
many promising results obtained under the LoS assumption are misleading about the actual RIS gain provided at random UE locations. Nevertheless, there are specific regions with a relatively high LoS probability where an SE gain can be achieved over point-to-point communications. In fact, the SE gain ranges from 80\% to 120\% for some RIS-deployment locations. Based on these findings, in mobile scenarios, it is evident that the RIS should be placed near the transmitter or receiver to ensure at least one LoS path (probability of one) for that link. Besides the FWA scenario, we have identified two additional mobile communication use cases where an RIS can make a great difference. These additional two use cases are shown in Fig.\ref{fig:RIS_placements}.b and c.

\textbf{Use Case 2: Enhanced Capacity for Distributed Users within the Cell} by positioning the RIS near the base station (transmitter) as illustrated in Fig.\ref{fig:RIS_placements}.a. If the RIS is situated in the always-LoS region of the transmitter-RIS link (e.g., $d_{2D}<18$\,m, which gives a LoS probability of 1 \cite{ETSI}), 
the transmitter and RIS can work in tandem, 
effectively creating a large massive MIMO array with a high beamforming gain thanks to the large RIS area. This scenario supports randomness for the UE locations within the cell since we guarantee an LoS link between the transmitter and RIS that mimics the left \textit{cyan-colored} region in Fig.~\ref{fig:RIS_placements}.a. Therefore, a performance increase is provided for all UEs in the cell. Indeed, those random UEs who are close to the RIS achieve significant SE improvements thanks to having LoS links for both hops.  To demonstrate the performance gain in this use case, Fig.~\ref{fig:cdf_vs_rate_incell}.a presents the cumulative distribution function (CDF) of the SE with the RIS located within the LoS region of the transmitting base station. The CDF is generated by considering 2000 user locations uniformly distributed within the cell. We notice that RIS-assisted communication enhances the CDF (moves it to the right) only when considering the static path. The improvements are around 2 bit/s/Hz throughout the entire CDF curves. This again proves the necessity of considering the static path in RIS-assisted point-to-point communication. Even if it is relatively weak, it has vital impact on the performance and cannot be neglected, as commonly done in the RIS literature.


Another prospective deployment scenario is related to \textbf{Use Case 3: Enhanced Capacity for a Region of Interest within the Cell}, as shown in Fig. \ref{fig:RIS_placements}.c, where the RIS is deployed close to the intended user locations (a region of interest). In this configuration, the LoS path is ensured between RIS and receiver for a region of interest to which the static path is likely to be nLoS. This region might be a blind spot within the cell coverage or an area where users typically need high data-rate applications like video streaming. Mobile network operators (MNOs) can identify these regions and deploy a nearby RIS. Placing an RIS near such a region is reminiscent of deploying a large antenna array that picks up signals from the base station and relays them to the intended users.


To assess the prospects of this use case, in Fig.~\ref{fig:cdf_vs_rate_incell}.b, we present the CDF of the SE at 200 user locations randomly distributed within the region of interest. As observed, the SE performance significantly improves when RIS-assisted communication is used with a static path. This gain is more pronounced at $7.8$\,GHz than at  $15$\,GHz, but one should remember that we consider an equal number of RIS elements in both cases so the physical dimensions are smaller at $15$\,GHz. Similar results are noted for RIS-assisted communication without a static path. This again highlights that the RIS is more advantageous with a static path, while the limitations are stricter without a static path. Nevertheless, it is obvious that RIS-assisted communication provides substantial gain when the RIS is located near the receiver (e.g., Use Case 3) rather than when it is near the transmitter (e.g., Use Case 2). Once RIS is near the receiver, the CDF curves are shifted to the right significantly (see Fig.~\ref{fig:cdf_vs_rate_incell}). This is achieved by the massive channel hardening effect at the receiver side. Note that Use Case 3 enables enhanced performance for a region of interest (e.g., a former blind spot). No performance gain can be guaranteed outside that region since the LoS probability decreases. Therefore, it does not support UE mobility, but if there is a limited number of blind spots in a cell, one RIS can be deployed for each of them.
This will evidently introduce deployment complexity and cost for MNOs, which must be carefully considered beforehand.

For both potential use cases in mobile communications, one might argue that implementing larger antenna arrays could be more beneficial than installing multiple RISs for each region or having an RIS near the base station. This point has been discussed in detail in \cite{RIS_power_scale}. Nonetheless, a significant advantage of using RIS is its energy-saving mechanism since it is a passive reflecting surface, and practical implementations \cite{RIS_implementation} show that it does not consume much power to operate. Conversely, massive MIMO systems can be power-intensive because they require many RF chains at the transceivers. A contender for Use Case 3 is to deploy small base stations that cover the blind spots, which is a viable alternative but requires additional components such as a backhaul infrastructure. 


\begin{figure*}[!t]
\centering
\fcolorbox{mycolor}{mycolor}{ 
\subfloat[Use Case 4: Increased Reliability at the Cell Edge]{\includegraphics[width=0.33\textwidth, height=4.6cm]{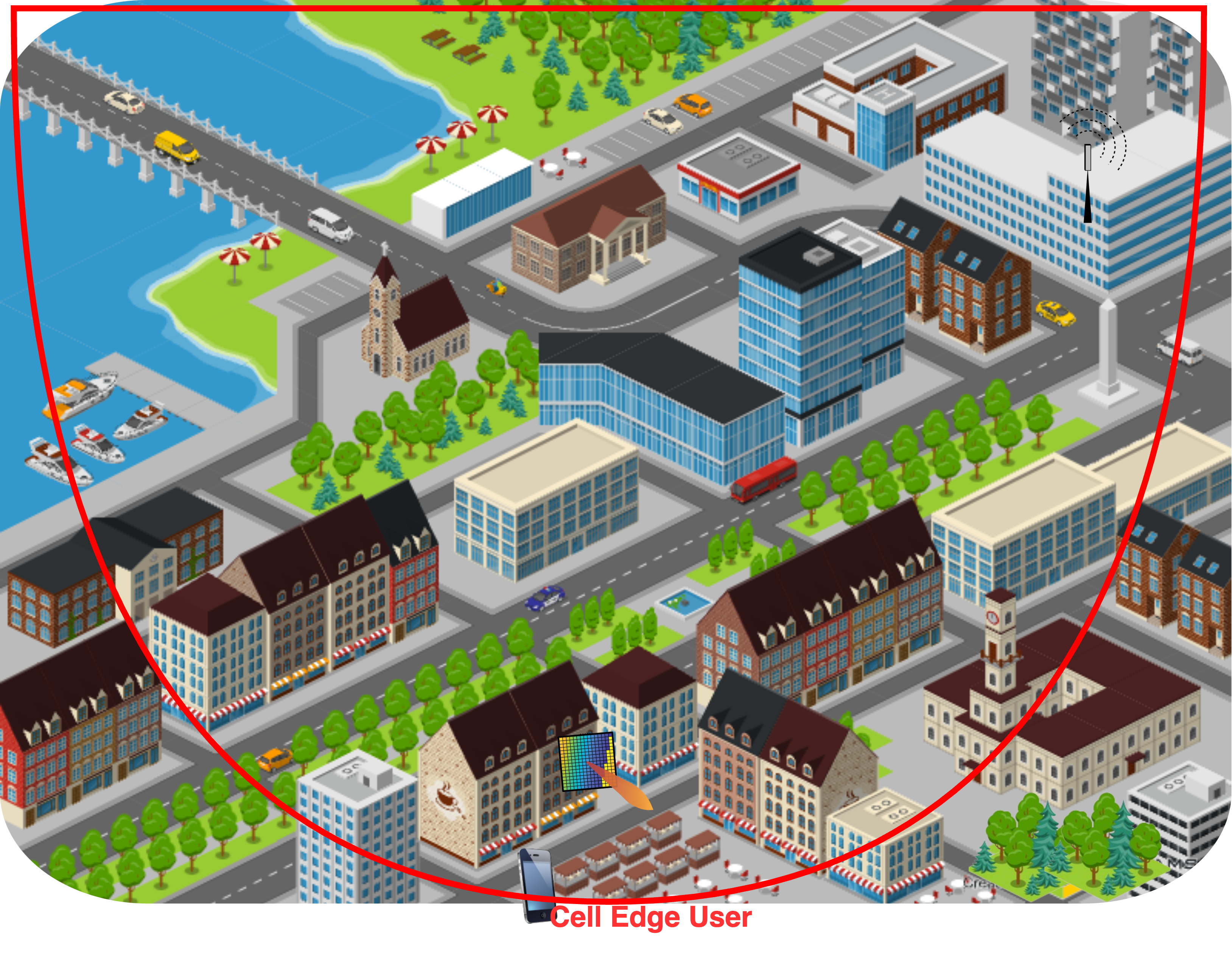}
}
\centering
\subfloat[Reliability vs Transmit Power]{\includegraphics[width=0.33\textwidth]{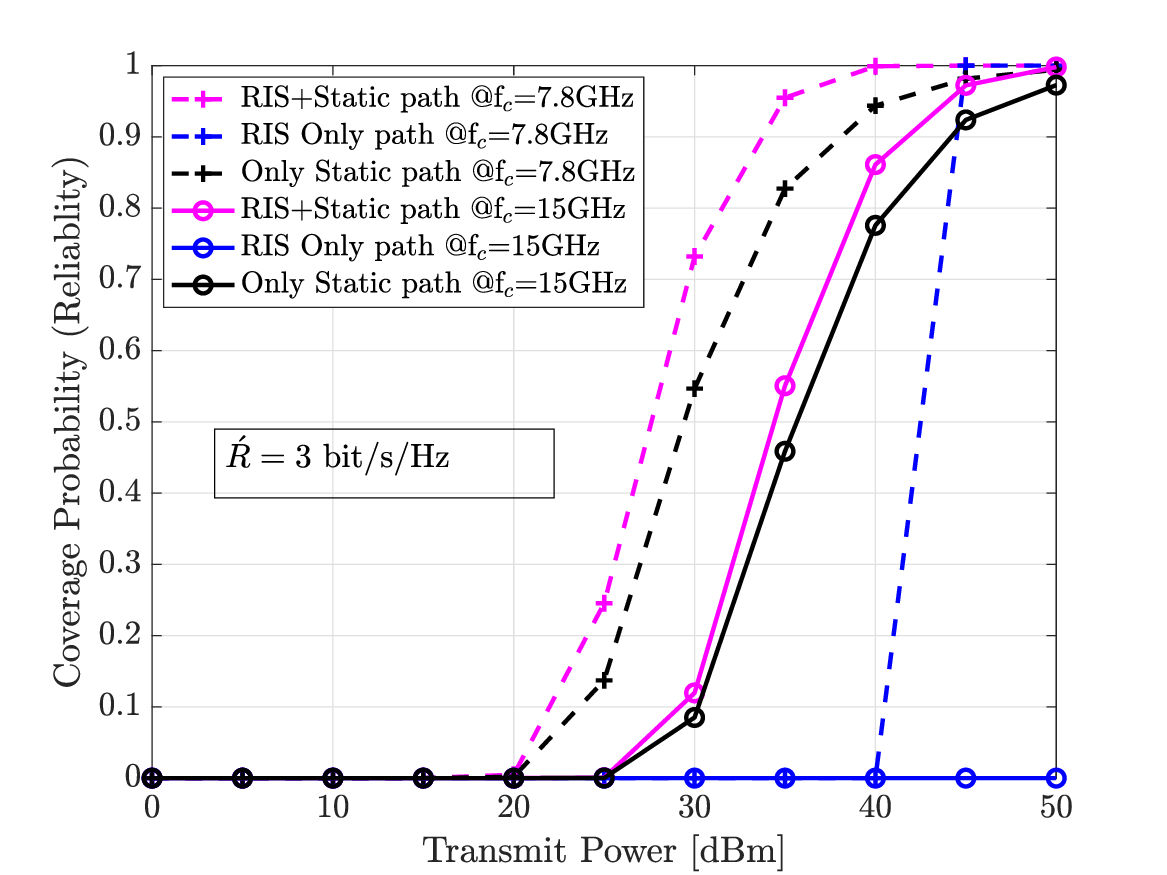}
}%
\centering
\subfloat[Reliability vs QoS Requirement]{\includegraphics[width=0.33\textwidth]{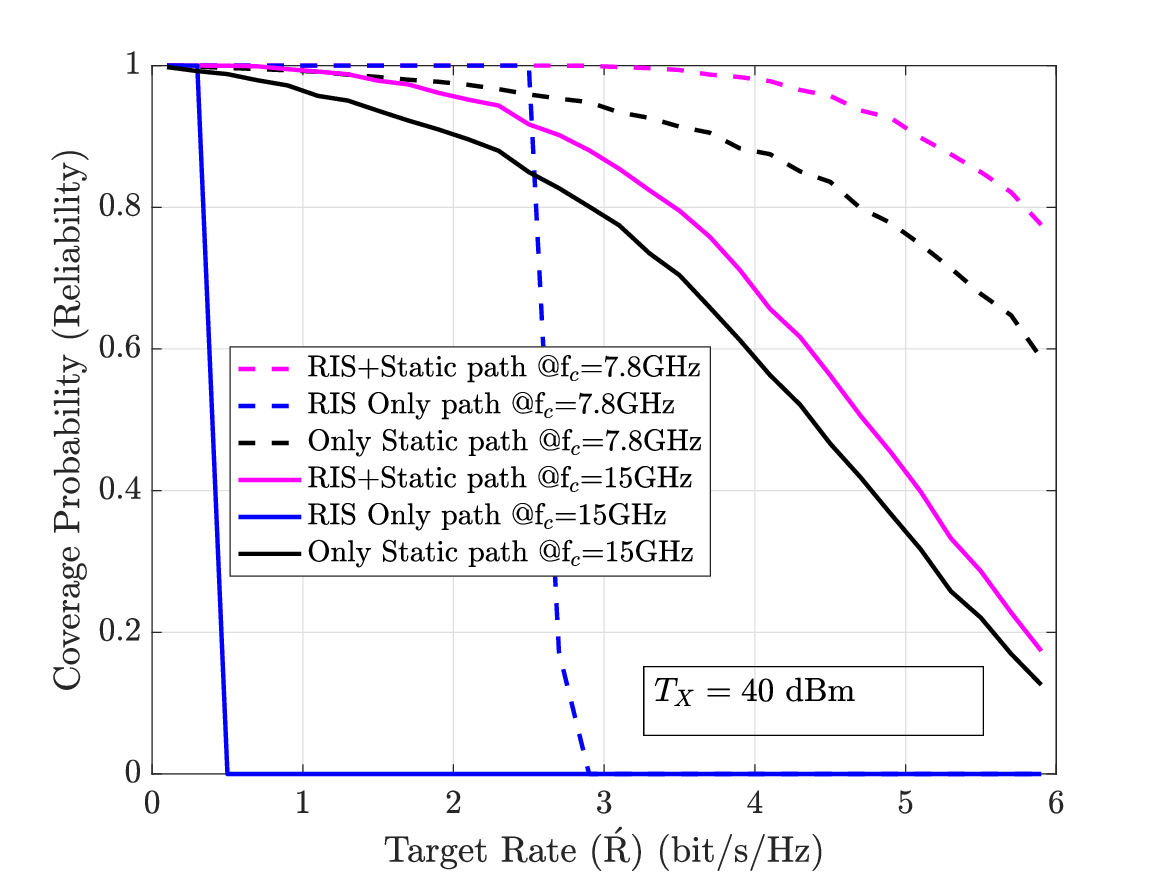}
}
}
\caption{Use Case 4: Increased reliability at the cell edge. The figures at the right-hand-side present coverage probability (reliability) performance comparisons of the cell-edge user when a RIS is deployed nearby (LoS region) the cell-edge.}
\label{fig:coverage_vs_transmit_power}
\end{figure*}

\subsection*{Increased Reliability at the Cell Edge}

In this section, we examine the network reliability performance for UEs at the cell edge. Fig.~\ref{fig:RIS_placements}.a showed that having an RIS node close to the users can enhance the average SE. However, network reliability is defined as the ability to maintain this SE over time. In other words, it is the probability of meeting the quality-of-service (QoS) requirements for the nodes. At the cell edge, without an RIS, users may fail to meet this requirement most of the time, even if the average achievable SE is significantly higher than the QoS requirement. This is due to varying channel conditions over time, but RIS can improve these conditions since many different fading realizations are obtained over its large surface.
We define this scenario as \textbf{Use Case 4: Improved QoS at the Cell Edge} as illustrated in Fig.\ref{fig:coverage_vs_transmit_power}.a. Suppose we place RIS nodes near the cell edges to enhance reliability for cell-edge users. 
There are two main benefits of doing this compared to increasing the transmit power: we are not causing extra inter-cell interference in an uncontrolled manner, and we alleviate small-scale fading by creating channel hardening \cite{Bjornson2021a}.
To demonstrate this, Fig.~\ref{fig:coverage_vs_transmit_power}.b presents the coverage probability (reliability) for a cell-edge user with a nearby RIS node. The coverage probability is defined as the probability of meeting a given QoS requirement (i.e., $\mathrm{Pr}(\mathrm{SE}\geq\acute{R})$). We observe that RIS-assisted communication with a static path consistently provides higher reliability than situations with only a static path. For example, in the same setup, Fig. \ref{fig:RIS_placements}.a shows that the point-to-point communication benchmark achieves an average SE of approximately 3\,bit/s/Hz at $T_X=30$\,dBm. However, Fig.~\ref{fig:coverage_vs_transmit_power}.b indicates that 45\% of the time (almost half), it fails to meet the QoS requirement of $\acute{R}=3$\,bit/s/Hz at $f_c=7.8$\,GHz. Conversely, with the aid of an RIS, the network reliability can increase to 75\%, guaranteeing the QoS requirement for the cell-edge user. The benchmark point-to-point communication requires at least $5$\,dB more transmit power to have similar reliability. When the transmit power is set to  $T_X=40$\,dBm, we see that RIS-assisted communication provides full reliability (i.e., a probability of 1) whereas the point-to-point communication can not guarantee $\acute{R}=3$\,bit/s/Hz at the 5\% of time although the average SE for the point-to-point communication will be around $\sim$ 6\,bit/s/Hz. However, if only the RIS path is used, significantly more transmit power (e.g., $\sim 45$\ dBm) is required to meet the QoS requirements.  In all scenarios, when $f_c=15$\,GHz with $\acute{R}=3$\,bit/s/Hz QoS requirement, the reliability values get worse due to severe path-loss and $N=2000$ elements cannot compensate for it. However, the RIS-only communication cannot provide this QoS requirement at all. This supports our assertion that RIS is more advantageous when a static path is also in use, contrary to much of the RIS literature.

To further investigate the performance of the cell-edge user, Fig.~\ref{fig:coverage_vs_transmit_power}.c presents how the coverage probability depends on the QoS requirements for a given transmit power. We observe that RIS-assisted communication with a static path offers up to 70\% higher QoS for the same reliability compared to point-to-point communication alone. Interestingly, unlike previous observations throughout the paper, RIS-only communication can outperform point-to-point communication if the QoS requirement is relatively low, guaranteeing 100\% coverage probability up to $\acute{R}=2.4$\,bit/s/Hz.
This is because the channel hardening effect is achieved by passive beamforming using many RIS elements \cite{Bjornson2021a}. Up until now in this paper, this is the only use case where RIS-only communication can outperform the point-to-point communication benchmark. However, when the rate requirement exceeds $\acute{R}=2.4$\,bit/s/Hz, the coverage probability of RIS-only communication rapidly drops to zero. Therefore, for certain QoS requirements, increasing reliability at the cell edge is the only use case where RIS-only communication provides a significant gain.


\section*{How Large Surfaces Are Required?}
In the previous sections, we utilized $N=2000$ elements for each scenario and found that RIS-only communication generally does not surpass point-to-point communication, particularly not at $f_c=15$\,GHz. In this section, our aim is to determine the minimum number of RIS elements required for RIS-only communication to match the performance of point-to-point communication in each use case. For the FWA use case, we can compute the minimum requirements, and based on the path loss definitions in \cite{ETSI} and through some mathematical manipulations, we derive that $N\geq \kappa f_c \sqrt[4]{d_{3D}}$ where $40<\kappa<45$ is a correction coefficient and $d_{3D}$ is the 3D distance between the transmitter and receiver (assuming the RIS is positioned in the middle, which is the most reasonable scenario in FWA to achieve LoS). However, this is not mathematically feasible for the other three use cases since the user positions vary randomly within the entire cell or a region of interest. Nevertheless, we can assert that the frequency will have a linear effect on the number of required RIS elements, and the distance between each node will have a nonlinear impact.

\begin{figure}[!t]
\centering
\fcolorbox{mycolor}{mycolor}{
\includegraphics[width=0.45\textwidth]{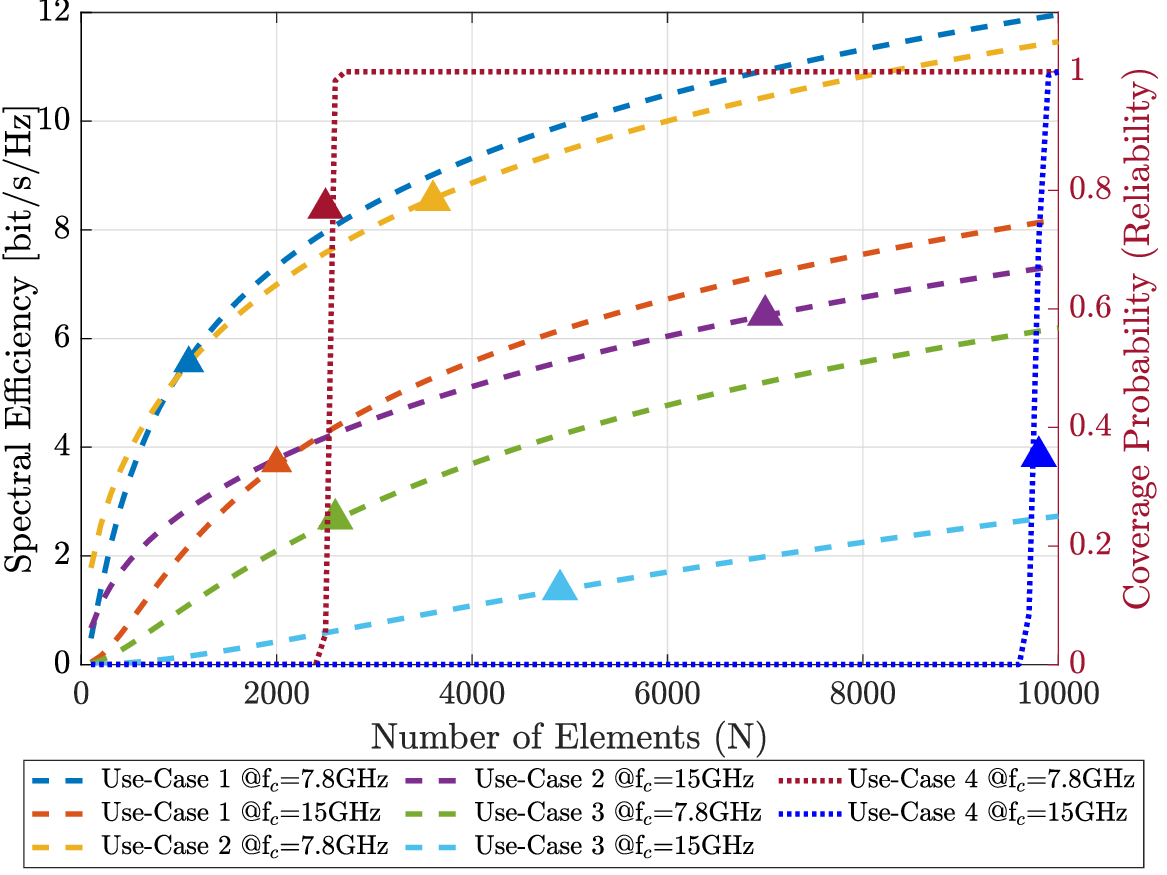}%
}
\caption{Performance comparisons of all use cases with respect to number of elements ($N$). Transmit power is $T_X=30$\,dBm. SE performance is given on the left $y$-axis for the first three use cases, and coverage probability is given for Use Case 4 on the right $y$-axis as a performance metric. $\acute{R}=2$\,bit/s/Hz is used for the QoS requirement in Use Case 4. For all use cases, the point-to-point communication benchmark performance points and the required number of elements for RIS-only communication to reach these benchmarks are illustrated with the markers. }
\label{fig:performance_vs_elements}
\end{figure}

To study these effects numerically, in Fig.~\ref{fig:performance_vs_elements}, we present the related performances of each use case regarding the number of elements at the RIS node. We considered exactly the same setups as in the previous simulations. We observe that the mathematical limit for FWA is valid, and only beyond that limit can RIS-assisted communication without a static path achieve the same performance as benchmarks. Furthermore, as predicted, with an increase in frequency, the number of RIS elements increases linearly, and from 7.8\,GHz to 15\,GHz, we clearly need to double the number of elements. In the previous discussion of Use Case 2 and Use Case 3, we noted that RIS-only communication could not achieve baseline performance with $N=2000$ elements, and now we see that in those scenarios, a minimum of $3000$-$7000$ elements are required. Additionally, with higher QoS requirements, RIS-only communication can achieve the same reliability threshold of only static path only if around 10\,000 RIS elements are needed when $f_c=15$\,GHz. A simple calculation shows that the 1-4\,m$^2$ surfaces are needed, which may not be feasible close to users but possibly at building facades.

\section*{Further Challenges and Open Research Problems}

We noticed in the previous section that physically large surfaces are often required to leverage the benefits of RIS in the upper mid-band. In practice, this presents several operational and technical challenges.

\textbf{Deployment of RIS and Site Rental Costs:} Firstly, the placement of large RIS
can be a showstopper for MNOs, especially in densely populated urban areas. It is not only difficult to find suitable deployment locations, but it also likely incurs high costs due to site rentals. 
MNOs may not find such deployments financially viable unless an RIS can be hidden behind other equipment such as commercial billboards. 

\textbf{Control of RIS and Interference Management:} An RIS has the ability to manipulate signals and redirect them in desired directions for MNOs. However, after reflecting off the RIS, the RF waves will again reflect from the wireless medium and scatterers, eventually causing interference. Additionally, in commercial systems, multiple MNOs operate in adjacent bands, and current RIS technology cannot filter out only the band of a single MNO but will affect all MNOs. 
Hence, if one MNO deploys an RIS and operates it according to one of the aforementioned use cases, the channel conditions in the adjacent MNOs' bands will also be modified. This leads to extra small-scale fading in the best case and inter-operator pilot contamination in the worst case \cite{RIS_pilot_contamination}.
This raises questions about whether each operator can have its own RISs or if all RISs must be shared and jointly operated by all MNOs. Since the upper mid-band is also adjacent to other sensitive systems, such as satellite links, the impact of RIS on coexistence must be carefully studied.

\textbf{Channel Estimation and Phase-Shift Optimization with Large RIS:}
In this paper, we have assumed that perfect channel state information (CSI) is always available.
However, estimating 10\,000 channel coefficients is challenging, not only from a signal processing perspective but also in terms of the immense pilot load. The latter problem can potentially be addressed by exploiting known channel characteristics, such as the existence of LoS paths whose angles can be estimated \cite{Haghshenas2024a}.
The use of such algorithms in practical environments with hardware imperfections and multipath components is yet to be demonstrated.
Furthermore, our results were obtained using the optimal phase shifts, which require infinite-resolution controllers at the RIS. Although near-optimal performance can be achieved with lower resolution, 
implementing a control unit for huge RIS element numbers can be very hard and costly.

\textbf{Near-Field Effects:} In this paper, we adopted the plane-wave assumption, assuming all nodes are located in each others' far-field regions. However, with large RIS sizes in the upper-mid band, radiative near-field effects will be inevitable when the users are close to the RIS (e.g., in Use Cases 3 and 4). In these situations, we end up with spherical waves with non-linear amplitude and phase variations across the RIS \cite{RIS_power_scale}. The simulation results in this paper can be seen as upper bounds since the radiative near-field effects can lead to power losses. Nevertheless, as shown in \cite{9723331}, with proper beamfocusing/phase optimization, this performance loss can be minimized. The near-field effect can also increase the MIMO rank, which can improve the performance. It is essential to develop an industry-accepted near-field-compliant channel model for RIS-assisted communication so performance can be properly evaluated and algorithms designed to capture all essential effects, there are more yet to be explored.  

\section*{Conclusions}

In this paper, we have studied RIS deployments in the upper mid-band to answer whether RIS can provide substantial gains or mostly cause technical and operational complexity (i.e., pain) with a marginal gain. 
First, we have demonstrated that a LoS link is crucial for RIS deployments in upper mid-band systems. Without it, RIS might be painful by introducing the need for highly complex algorithms (i.e., channel estimation and phase-shift design for large elements) in return for a subtle gain. Based on this observation, we have identified four prospective use cases for RIS deployments where an RIS node significantly improves performance.  Within these four use cases, enhancing the range and capacity for FWA has a clear commercial value since more customers can be accommodated per base station.
There are multiple ways to deploy RISs to improve the SE of mobile users, but when considering realistic channel models with a non-zero static path and uncertain LoS to the RIS, the gains become rather modest. The required physical RIS size can also be a showstopper. Nevertheless, an RIS can greatly improve performance, particularly when located near the end nodes. This could either refer to a blind spot with multiple users or a cell-edge user.  In these cases, RIS-assisted communication guarantees a substantial increase in SE  for a blind spot or greatly enhanced network reliability for cell-edge users by combating fading through channel hardening. These might be the most promising use cases.
Many practical problems must be solved before any commercial RIS deployments can begin, including the coexistence between MNOs and other services.


 

%
\ifCLASSOPTIONcaptionsoff
  \newpage
\fi
\bibliographystyle{IEEEtran}

\bibliography{ref}
%
\end{document}